\documentclass[aps,twocolumn,showpacs,amsmath]{revtex4-1}

\usepackage{dcolumn}    
\usepackage{bm}         
\usepackage{ifpdf}
\usepackage{amssymb,lineno,amsfonts}
\usepackage{graphicx}   
\usepackage{bbm}
\usepackage{mathrsfs}
\usepackage{upgreek}
\usepackage{mathtools}
\usepackage{epstopdf}
\usepackage{setspace}
\usepackage{hyperref}
\usepackage[matrix,frame,arrow]{xypic}
\usepackage{rotating}
\usepackage{float}

\begin{document}
\title{Characterization of hyperfine interaction between an NV electron spin and a first shell  $^{13}$C nuclear spin in diamond}

\author{K. Rama Koteswara Rao and Dieter Suter}
\affiliation{Fakult{\"a}t Physik, Technische Universit{\"a}t Dortmund}
\date{\today}


\begin{abstract}
{
The Nitrogen-Vacancy (NV) center in diamond has attractive properties for a number of quantum technologies
that rely on the spin angular momentum of the electron and the nuclei adjacent to the center.
The nucleus with the strongest interaction is the $^{13}$C nuclear spin of the first shell. 
Using this degree of freedom effectively hinges on precise data on the hyperfine interaction between the electronic and the nuclear spin.
Here, we present detailed experimental data on this interaction, together with an analysis that yields all parameters of the hyperfine 
tensor, as well as its orientation with respect to the atomic structure of the center.
}
\end{abstract}

\maketitle


Nitrogen-Vacancy (NV) centers in diamond have interesting properties for applications in room-temperature metrology, spectroscopy, 
and Quantum Information Processing (QIP)  \cite{Lukin2007Sci, JW2008Nat, JW2008Sci, JW2011NatPhy, JW2013Sci, JW2014Nat}. 
Nuclear spins, coupled by hyperfine interaction to the electron spin of the NV-center, are important for many of these applications 
\cite{Chil2006Sci,Lukin2007Sci,JW2008Sci,JW2010Sci,Chil2011, JW2014Nat, Han2014NatNano, Alvarez2015, Fuchs2011, Lukin2012Sci}.
For example, in QIP applications, the nuclear spins can hold quantum information \cite{Fuchs2011, Lukin2012Sci}, 
serving as part of a quantum register \cite{Lukin2007Sci}.
Accurate knowledge of the hyperfine interaction is necessary, e.g., for designing precise and fast control sequences for the nuclear spins \cite{ChilPRA2009, PaolaPRB2015}. 
With a known Hamiltonian, control sequences can be tailored by optimal control  techniques to dramatically improve the speed and precision 
of multi-qubit gates \cite{Khaneja2005, Khaneja2007, ChouPRA2015, DuNatCom2015}.

The $^{13}$C nuclear spin of the first coordination shell is a good choice for a qubit due to its large hyperfine coupling to the electronic spin of the NV center, 
which can be used to implement fast gate operations \cite{Jelez2004, JW2008Sci, JW2009PRB} or high-speed quantum memories \cite{Shim2013}. 
Full exploitation of this potential requires accurate knowledge of the hyperfine interaction, including the anisotropic (tensor) components.
The interaction tensor has been calculated by  DFT  \cite{Horodecki2004,Gali2008, Nizovtsev2010, GaliPRB2013},
but only a limited number of  experimental studies of this hyperfine interaction exist to date \cite{vanWyk1978, Felton2009, ShimArXiv}.
In this work, we present a detailed analysis of this hyperfine interaction.
The experiments were carried out on single NV centers of a diamond crystal with a natural abundance of $^{13}$C and a
nitrogen concentration of $<5$ ppb using a home-built confocal microscope and microwave electronics for excitation \cite{Shim2013}. 

\begin{figure}[h]
	\centering
	\includegraphics[width=8.8cm]{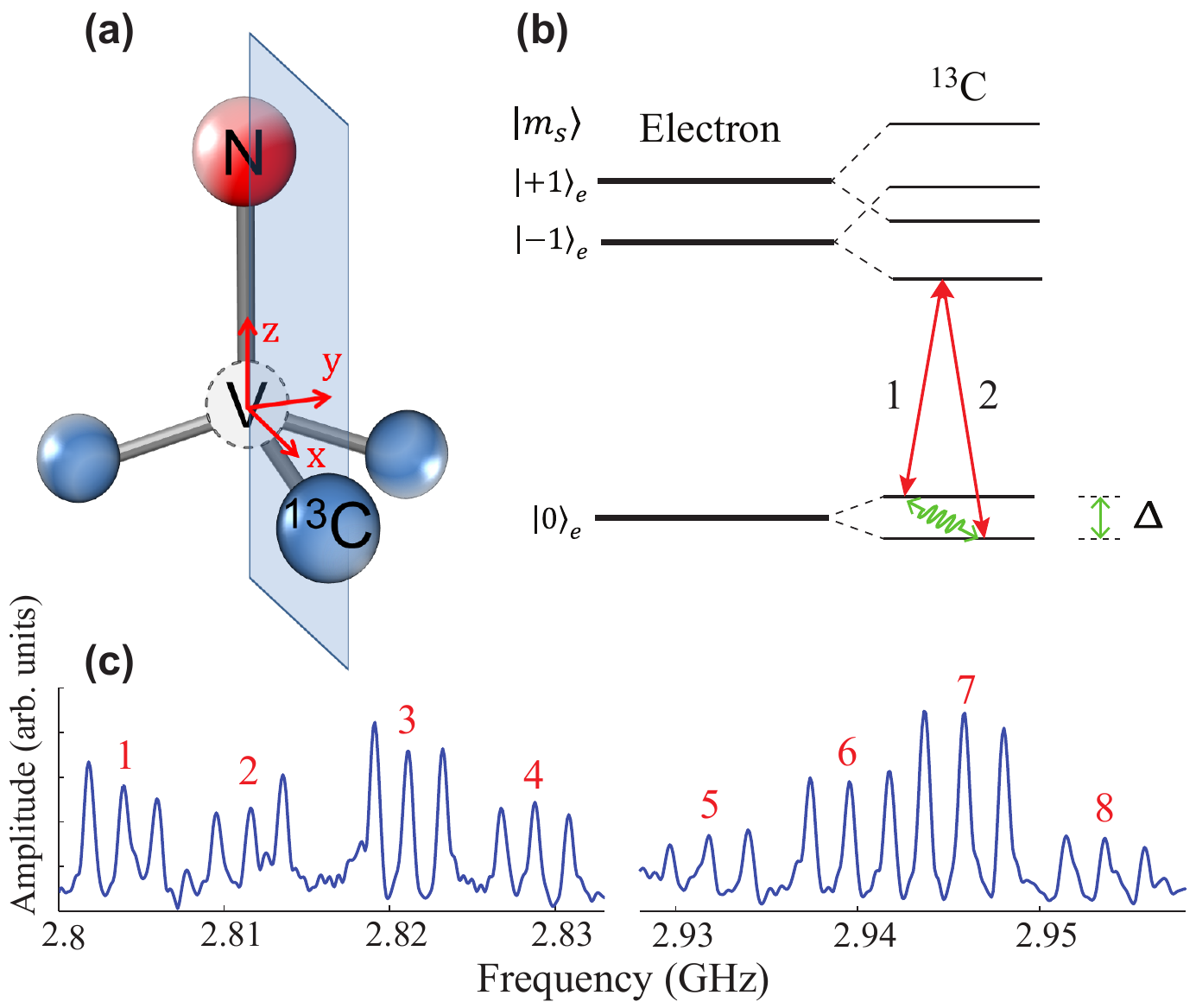}
	\caption{Symmetry of the system along with the energy level diagram and spectra. 
	(a) Structure of the NV center in diamond with a first shell $^{13}$C. 
	The mirror plane and the $x$, $y$, and $z$- axes of the NV frame are also shown. 
	(b) Energy level diagram considering only the electron and $^{13}$C nuclear spins. 
	The red arrows and green wave represent electron- and nuclear spin transitions, respectively. 
	(c) Experimental ESR spectrum for a particular orientation of the magnetic field vector. 
	The spectrum consists of 24 resonance lines due to the interaction with the $^{14}$N nuclear spin.}
	\label{struc} 
\end{figure}

The presence of a  nuclear spin in the first coordination shell reduces the symmetry of the NV center from $\mathrm{C_{3V}}$ to $\mathrm{C_S}$, 
a single mirror plane. 
This symmetry plane passes through the NV symmetry axis and the $^{13}$C nuclear spin as illustrated in Fig.\,\ref{struc}(a). 
The NV symmetry axis  defines the $z$-axis of the NV frame of reference,
the $x$-axis lies in the symmetry plane of the center, and the $y$-axis is perpendicular to both of them. 
Due to the symmetry of the system, only those elements of the hyperfine tensor  that are invariant with respect to the inversion of the $y$-coordinate can be non-zero. 
Hence, the hyperfine  Hamiltonian in the NV frame of reference can be written as
\begin{equation}
{\cal H}_{\textrm{hf}}=A_{zz} S_z I_z + A_{xx} S_x I_x + A_{yy} S_y I_y + A_{xz} (S_x I_z+ S_z I_x).
\label{eq:HypHamiltonian}
\end{equation}
Here, $S_\alpha$ and $I_\alpha$ represent components of the electron and nuclear spin angular momenta, respectively, 
and $A_{\alpha\beta}$  the components  of the hyperfine tensor.
The full Hamiltonian of the system consisting of an NV electron spin and  a $^{13}$C nuclear spin can be written as
\begin{equation}
{\cal H}=D S_z^2 + \gamma_e \mathbf{B}\cdot \mathbf{S} + \gamma_n \mathbf{B}\cdot \mathbf{I} + {\cal H}_{\textrm{hf}}.
\label{eq:Hamiltonian}
\end{equation}
Here, $D$ represents the zero-field splitting of the electron spin, $\gamma_e$ and $\gamma_n$ are the gyromagnetic ratios of the electron and nuclear spins respectively, 
 ${\mathbf{B}} = B(\sin \theta \cos \phi, \ \sin \theta \sin \phi, \ \cos \theta)$ represents the magnetic field vector and $\theta$ and $\phi$  its polar and azimuthal angles in the NV frame of reference.
For typical magnetic fields, the nuclear Zeeman interaction is much smaller than the other components of the Hamiltonian 
and the quantization axis of the nuclear spin is dominated by the hyperfine interaction.
In the present analysis, we ignore the hyperfine interaction of the $^{14}$N nuclear spin, which is small compared to the hyperfine interaction under investigation. 
Fig.\,\ref{struc}(b) shows the energy level diagram of the system.
There are  eight possible transitions of the electron spin from the two $m_S=0$ states to the four  $m_S=\pm1$ states.
The first two of these transitions are shown by red arrows, while the green wave marks the nuclear spin transition within the  $m_S=0$ multiplet. 
Fig.\,\ref{struc}(c) shows a typical experimental Electron Spin Resonance (ESR) spectrum containing 24 resonance lines. 
The additional factor of 3 is due to the hyperfine interaction with the $^{14}$N nuclear spin, which we neglect in the present context.

We start the determination of the hyperfine tensor by measuring ESR spectra for different orientations of the magnetic field vector 
and then fitting the resonance frequencies to obtain all relevant Hamiltonian parameters, including the orientation of the NV center with respect to the laboratory frame. 
As a first step we determined the orientation of the NV-axis of the center, which corresponds to the $z$-axis of the Hamiltonian of Eq. (\ref{eq:Hamiltonian}).
For this purpose, we rotated the magnetic field with a fixed magnitude around two orthogonal rotation axes crossing at the NV center.
ESR spectra were measured as Fourier transforms of Ramsey fringes for nine non-coplanar orientations of the magnetic field.
The transition frequencies of these spectra were fitted numerically to determine the orientation of the $z$-axis of the center.
This orientation was reconfirmed by also measuring spectra of a neighboring NV center with no  $^{13}$C nucleus in the first coordination shell.
The two data sets yielded the same direction for the $z$-axis. 
If the field is aligned with the $z$-axis, the splitting between the transitions $m_s$=0 $\leftrightarrow m_s$=$\pm$1 reaches a maximum.
The numerical fit also provided estimates of  the Hamiltonian parameters
$D\approx 2870.2$ MHz, $\gamma_e B \approx 63.3$ MHz, $\sqrt{A_{zz}^2+A_{xz}^2} \approx 131$ MHz,
which give the dominant contribution to the transition frequencies when $\mathbf{B} || z$. 

\begin{figure} [h]
	\centering
	\includegraphics[width=8.2cm]{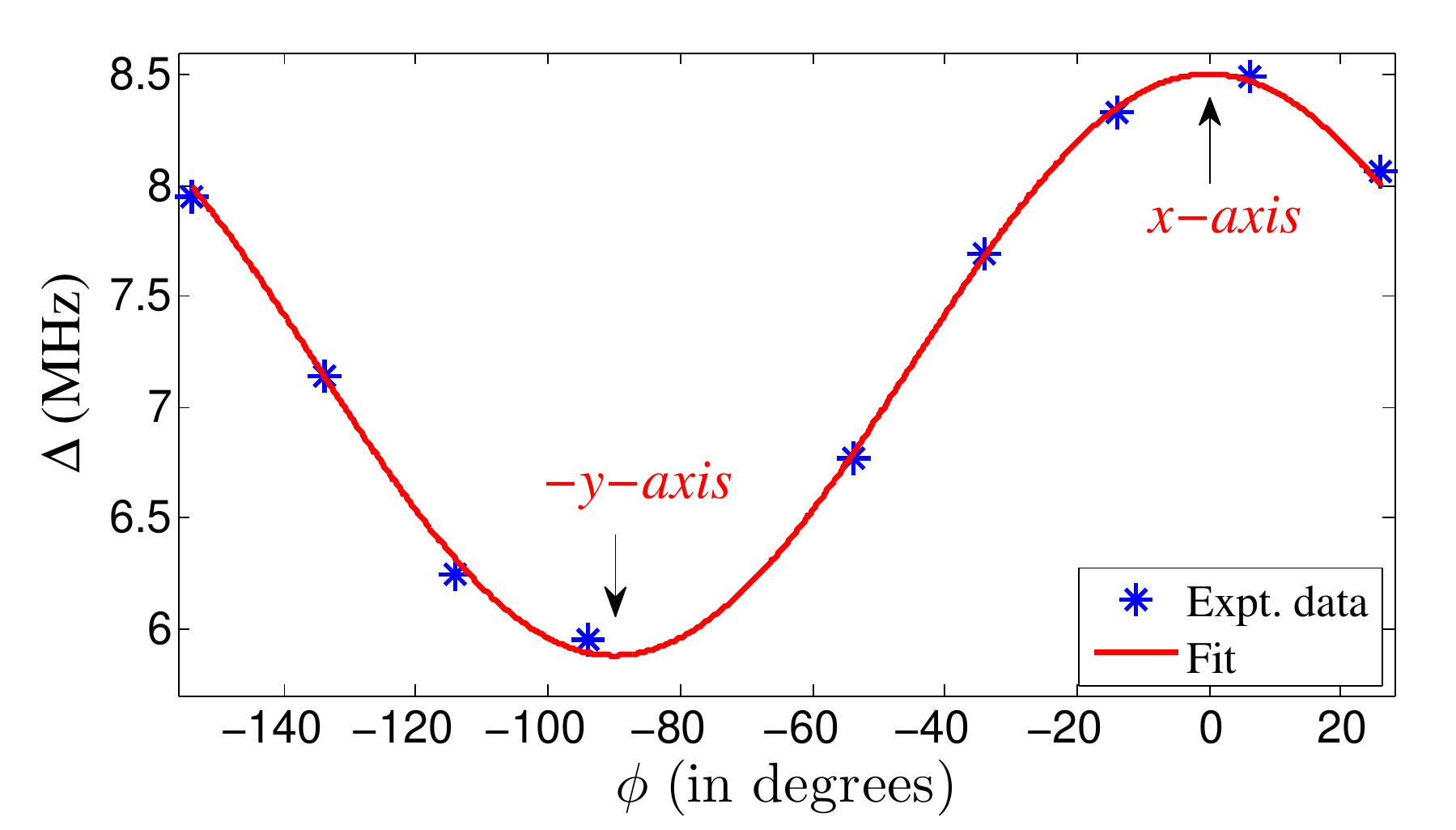}
	\caption{Zero-quantum transition frequencies for an azimuthal ($\phi$) rotation of the magnetic field for $\theta$=84.5$^{\circ}$. 
	The experimental data (blue stars) 
	were fitted to the equation $\kappa_1 \cos^2 \phi+\kappa_2 \sin^2 \phi$.
	The fit parameters are $\kappa_1$ = 8.5 MHz and $\kappa_2$ = 5.88 MHz.}
	\label{zqfig} 
\end{figure}

The $A_{xx}$ and $A_{yy}$ components of the hyperfine tensor contribute only in second order to the energies and spectral positions.
Accordingly, their uncertainty is quite large.
Their influence on the transition frequencies is maximised if the magnetic field is close to the $xy$-plane.
The highest precision is obtained by measuring the transitions with the smallest linewidth.
In the NV system, this is the nuclear spin transition between the $m_S=0, m_I=\pm1/2$ states. 
As a nuclear spin transition, the width of this resonance line is about one order of magnitude smaller ($\approx 60$ kHz) than that of the ESR transitions.
This transition can be excited not only by resonant radio-frequency pulses, but also by the Raman excitation scheme shown  in Fig.\,\ref{struc}(b):
A microwave pulse that drives the transitions from both $|m_S=0,m_I=\pm 1/2\rangle$ states to one of the $m_S=\pm1$ states creates nuclear spin
coherence and can also probe this nuclear spin coherence by converting it back into population of the $m_S = 0$ state.
The optimal duration of the microwave excitation pulse for the nuclear spin coherence is in general twice as long as that of the optimal pulse for the ESR transitions. 
Apart from the narrower linewidth, the nuclear spin transition can be identified in the ESR spectrum as a zero-quantum transition:
Its position does not change with the carrier frequency used for the excitation and readout pulses.
Using second-order perturbation theory, its frequency can be written as \cite{ShimArXiv} 
\begin{equation}
\Delta \approx \frac{2 | \gamma_e B \sin \theta|}{D}(\sqrt{A_{xx}^2+A_{xz}^2} \cos^2\phi + |A_{yy}| \sin^2\phi).
	\label{eq.ZQFreq} 
\end{equation}
Therefore, by measuring these zero-quantum frequencies for different orientations of the magnetic field for a fixed $\theta$,
it is possible to obtain estimates of the quantities $\sqrt{(A_{xx}^2+A_{xz}^2)}$ and $\lvert A_{yy} \rvert$.
Fig.\,\ref{zqfig} shows a graphical representation of the experimental data, measured for magnetic fields oriented at an angle $\theta = 84.5^\circ$ from the $z$-axis, 
together with the numerical fit.

From the fitted curve, we find that the parameters $\sqrt{(A_{xx}^2+A_{xz}^2)}$ and $\lvert A_{yy} \rvert$ 
must have the values $\approx$ 193.0 and 133.5 MHz.
The quantity $\sqrt{(A_{xx}^2+A_{xz}^2)}$ is larger (smaller) than $\lvert A_{yy} \rvert$ if the maximum (minimum) of the zero-quantum frequencies corresponds to the $x$-axis.
Combining these data with those for $D$, $\sqrt{(A_{zz}^2+A_{xz}^2)}$, and $\gamma_e B$, we obtain several possible parameter sets, 
with different signs, two different orientations for the $x$- and $y$-axes, and different ratios $A_{xz}/A_{zz}$.

To eliminate the remaining ambiguities and to determine the orientation of the $x$ and $y$-axes, 
it is important to use not only the transition frequencies, but also the transition amplitudes (dipole moments).
For this purpose, we measured the transition amplitudes of the spectral lines for different orientations of the magnetic field.
The experimental amplitudes depend on the orientation of the microwave magnetic field. 
Since absolute amplitudes are hard to measure, we determined ratios of transitions amplitudes for two different data sets:
First, we measured ratios of Rabi frequencies for pairs of transitions whose frequencies differ by the nuclear spin transition frequency 
discussed above. These transitions connect the two $m_S = 0$ states with the same $m_S = \pm1$ state.
In the second set of data, we compared the transition amplitudes of spectra in the $xy$-plane.
The first set of data indicated that the $x$-axis corresponds to the maximum of the zero-quantum frequency (see Fig. \ref{zqfig}).
This implies that  $\sqrt{(A_{xx}^2+A_{xz}^2)} \approx 193$ MHz and $\lvert A_{yy} \rvert \approx 133.5$ MHz. 
These data also indicate that if $A_{xx}$ and $A_{zz}$ have the same sign, the ratio $|A_{xz}/A_{zz}|$ must be $<0.3$.

\begin{figure}[h]
	\centering
	\includegraphics[width=8.4cm]{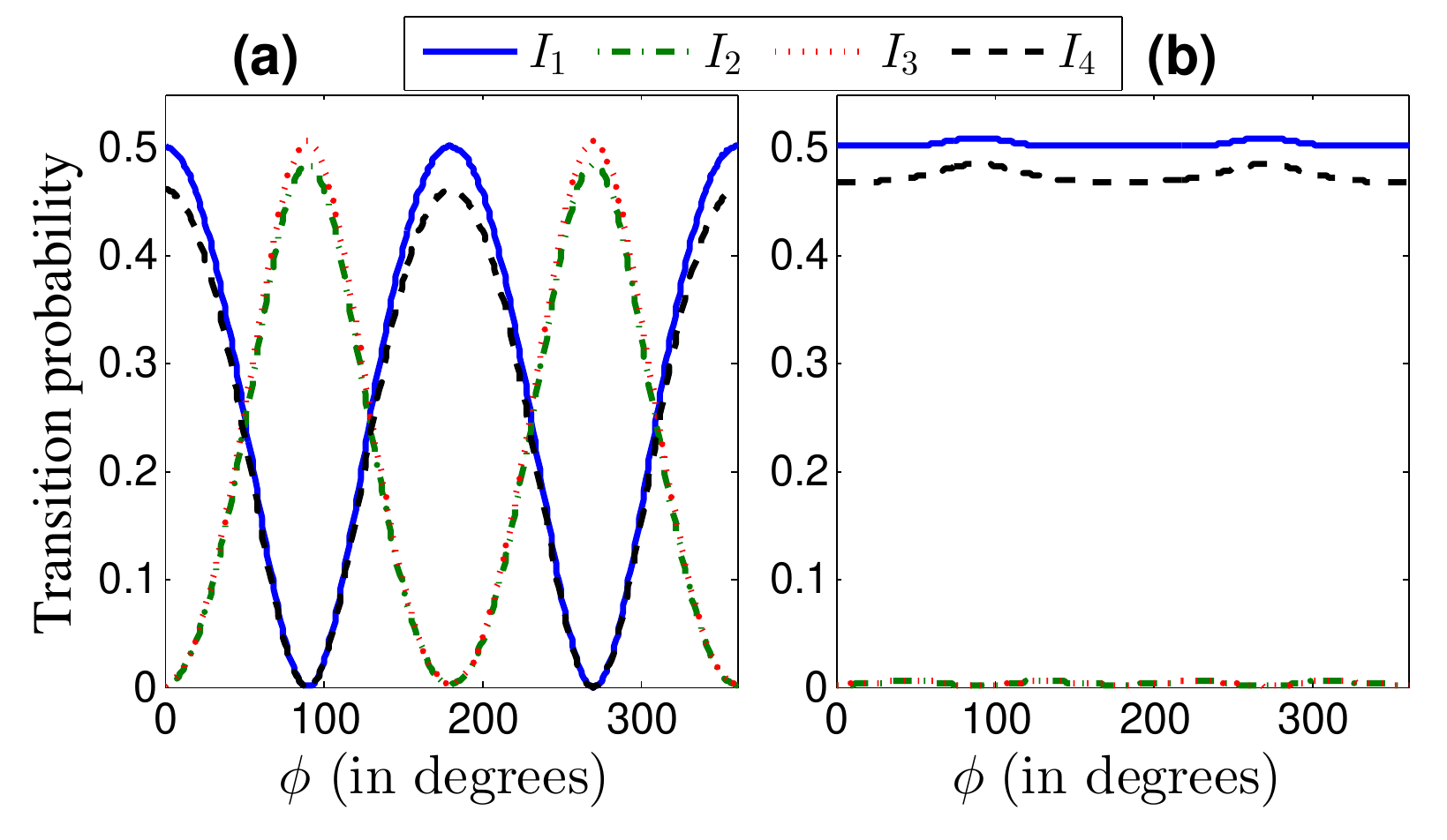}
	\caption{Transition probabilities of low frequency ESR lines ($\nu < 2870$ MHz) when $\mathbf{B}$ is oriented in the $xy$-plane.  
	(a) and (b) correspond to different sign combinations of the hyperfine parameters. In (a), $\det(A)$ $>$ 0 and in (b), $\det(A)$ $<$ 0.}
	\label{sq_sign} 
\end{figure}

\begin{figure}[h]
	\centering
	\includegraphics[width=8cm]{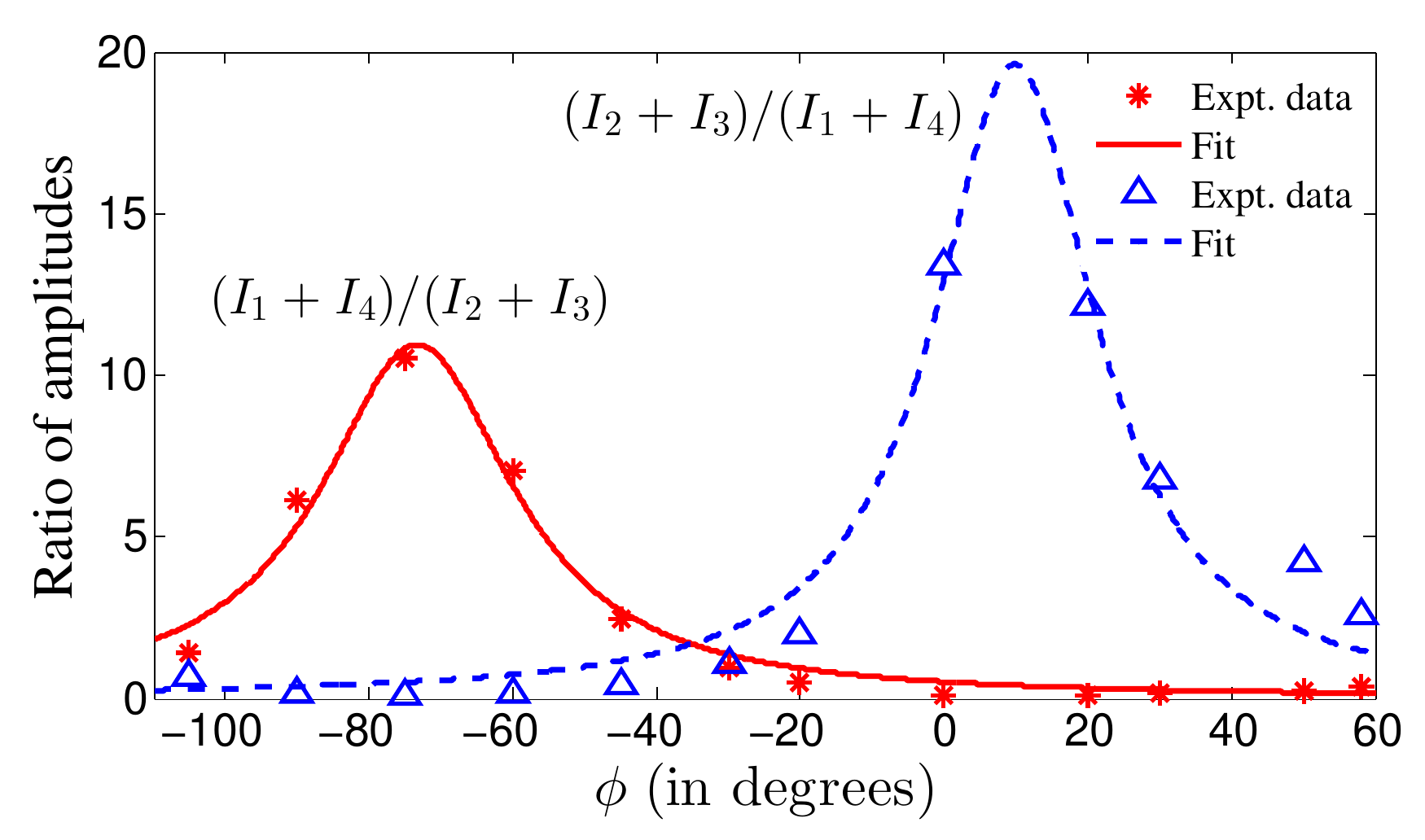}
	\caption{Ratios of amplitudes of the low-frequency spectral lines ($\nu < 2870$ MHz) in the $xy$-plane as a function of $\phi$. 
	The experimental data (red stars and blue triangles) were fitted to a lorentzian $a b/((\phi-\phi_1)^2+b^2)$  (red solid and blue dashed lines). 
	For the red solid line, the fit parameters are $a$=178.5, $b$=16.3, and $\phi_1$=-73.3$^\circ$ and for the blue dashed line $a$=270.5, $b$=13.8, and $\phi_1$=9.9$^\circ$. 
	}
	\label{sq_expt} 
\end{figure}

Fig. \ref{sq_sign} shows the numerically calculated transition probabilities  of the low-frequency spectral lines ($\nu < 2870$ MHz) 
as a function of the azimuthal angle $\phi$ of the magnetic field in the $xy$-plane.
Here, we observe two qualitatively different cases, depending on the relative signs of the hyperfine parameters.
Labeling the transition probabilities in ascending frequency order as $I_1 \dots I_4$, we observe strong variations
when $\det(A) > 0$, with $I_1$ and $I_4$ in phase and  $I_2$, $I_3$ shifted by $90^\circ$.
However, for $\det(A) < 0$, the amplitudes are almost constant, with $I_2$ and $I_3$ much smaller than $I_1$ and $I_4$.
As shown in Fig. \,\ref{sq_expt}, the experimental data show large variations of the transition probabilities, 
which is well compatible with the case $\det(A) > 0$  and excludes the case $\det(A) < 0$.
This allows us to disregard the parameter sets with $\det(A) < 0$ in the following.
Also, the difference in the position of the peak of the quantity $(I_1+I_4)/(I_2+I_3)$ in Figs. \ref{sq_sign} and \ref{sq_expt} allows us to determine 
the direction of the transverse component of the microwave field with respect to the $x$-axis of the NV frame.

From a combined fit of the available experimental data, we thus obtained values for all  components of the hyperfine tensor. 
Table \ref{resNVtab} lists the different parameter sets, in the NV frame of reference, that are compatible with the experimental data.
The different solutions all have the same principal components, except for the signs. 
The signs of $A_{xz}$ can be chosen positive or negative; the sign change corresponds to a $\pi$-rotation around the $z$-axis.

\setlength{\tabcolsep}{7pt}
\renewcommand{\arraystretch}{1.5}
\begin{table}[h]
	\begin{center}
		\begin{tabular}{| c | c |  c | c | c|}
			\hline
			Sol. \# & 1 & 2 & 3 & 4\\	
			\hline
			\hline	
			$A_{xx}$ (MHz) & 189.3  &  -189.3 & -163 & 163 \\
			\hline	
			$A_{yy}$ (MHz) & 128.4  & 128.4 & -128.4  & -128.4\\
			\hline	
			$A_{zz}$ (MHz) & 128.9 & -128.9 & 85.7 & -85.7\\
			\hline
			$A_{xz}$ (MHz) & $\pm$ 24.1 & $\mp$ 24.1 & $\mp$ 99.3 & $\pm$ 99.3\\
			\hline
		\end{tabular}
	\end{center}
	\caption{Hyperfine tensor components in the NV frame of reference.}
	\label{resNVtab}
\end{table}

It is useful to consider also the principal axis representation of the hyperfine tensor.
We  write $\overline{A}_{xx}$, $\overline{A}_{yy}$, and $\overline{A}_{zz}$ for the principal components of the hyperfine tensor
in increasing magnitude.
Table \ref{restab} lists the possible values for the principal components. 
All four parameter sets result in identical transition frequencies and amplitudes and are therefore experimentally indistinguishable.
The $y$-axes of the principal axis system (PAS) and the NV frame of reference coincide and the angle between the NV symmetry axis and the $z$-axis of the PAS is $\zeta$=109.3$^\circ$.
As a second-rank tensor, the hyperfine tensor is invariant under $\pi$-rotations around the principal axes.
Accordingly, orientations with the angles $-\zeta$ and $180^\circ \pm \zeta$ are equivalent solutions. 
All of these solutions are compatible with the experimental data.
Since DFT calculations  \cite{Gali2008, BudkerPRB2013} indicate that the largest component of the hyperfine tensor, which we write as $\overline{A}_{zz}$,
points in the direction of the $^{13}$C atom, the solution $\zeta$=109.3$^\circ$ appears to be the most meaningful one,
since it agrees very well with the theoretical value $109.5^\circ$ obtained from the geometry.

\setlength{\tabcolsep}{7pt}
\renewcommand{\arraystretch}{1.5}
\begin{table}[h]
	\begin{center}
		\begin{tabular}{| c | c |  c | c | c|}
			\hline
			Sol. \# & 1 & 2 & 3 & 4\\	
			\hline
			\hline	
			$\overline{A}_{xx}$ (MHz) & 120.5  &  -120.5 & -120.5 & 120.5 \\
			\hline	
			$\overline{A}_{yy}$ (MHz) & 128.4  & 128.4 & -128.4  & -128.4\\
			\hline	
			$\overline{A}_{zz}$ (MHz) & 197.8 & -197.8 & 197.8 & -197.8\\
			\hline
		\end{tabular}
	\end{center}
	\caption{Hyperfine tensor components in its PAS.}
	\label{restab}
\end{table}

The hyperfine tensor components determined here are in reasonable agreement with other values found in the literature. 
The values from the earlier ensemble EPR measurements \cite{Felton2009, BudkerPRB2013} are $\overline{A}_\parallel$=199.7 and $\overline{A}_\perp$=120.3 MHz, 
and those from the DFT calculations \cite{Gali2008, GaliPRB2013} are $\overline{A}_{xx}$=114.0, $\overline{A}_{yy}$=114.1, and $\overline{A}_{zz}$=198.4 MHz.
However, the earlier experimental studies assumed uni-axial symmetry of the hyperfine tensor and there was no information about the signs of the hyperfine components. 
Here, we measured the deviation from the uni-axial symmetry and found four equivalent sign combinations.

The parameter set determined in Ref. \cite{ShimArXiv}, which was given as $A_{xx}$=166.9, $A_{yy}$=122.9, $A_{zz}$=90.0, and $A_{xz}$=$-$90.3 MHz in the NV frame of reference, 
is comparable in magnitude with the solutions $3$ and $4$ of Table \ref{resNVtab}.
However, the signs of the parameters and the ratio $|A_{xz}/A_{zz}|$ together are not compatible with the measured ratios of Rabi frequencies of ESR transitions as discussed earlier.

In conclusion, we have performed a detailed analysis of the hyperfine interaction between an NV electron spin and a $^{13}$C nuclear spin of the first  shell. 
This analysis yielded accurately the hyperfine tensor and its PAS.
The present study will be helpful for implementing precise control operations in quantum registers containing  the first-shell $^{13}$C nuclear spin of the NV center. 
This nuclear spin is particularly attractive because of its strong coupling to the NV electron spin, 
which is necessary for implementing fast gate operations in hybrid quantum registers consisting of electron and nuclear spins.

We gratefully acknowledge experimental assistance from J. Zhang and useful discussions with F. D. Brand\~{a}o, J. H. Shim, and T. S. Mahesh. 
This work was supported by the DFG through grant Su 192/31-1.

\bibliography{bibNV1}

\end{document}